\documentclass[aip,jcp,reprint,groupedaddress,floatfix]{revtex4-1}

\usepackage{graphicx}
\usepackage[version=4]{mhchem}
\usepackage{microtype}
\usepackage{multirow}
\usepackage{physics}
\usepackage{siunitx}

\usepackage{hyperref}

\renewcommand{\vec}[1]{\mathbf{#1}}

\newcommand{\adj}{^\dagger}

\newcommand{\lmax}{\ell_\text{max}}

\newcommand{\qnlp}{\ell_\text{p}}
\newcommand{\qnm}{m}

\begin{document}

\title{
	Ground states of linear rotor chains via the density matrix renormalization group
}

\author{Dmitri Iouchtchenko}
\author{Pierre-Nicholas Roy}
\email{pnroy@uwaterloo.ca}
\affiliation{Department of Chemistry, University of Waterloo, Waterloo, Ontario, N2L 3G1, Canada}

\begin{abstract}
In recent years, experimental techniques have enabled the creation of endofullerene peapod nanomolecular assemblies.
It was previously suggested that the rotor model resulting from the placement of dipolar linear rotors in one-dimensional lattices at low temperature has a transition between ordered and disordered phases.
We use the density matrix renormalization group (DMRG) to compute ground states of chains of up to 50 rotors and provide further evidence of the phase transition in the form of a diverging entanglement entropy.
We also propose two methods and present some first steps towards rotational spectra of such nanomolecular assemblies using DMRG.
The present work showcases the power of DMRG in this new context of interacting molecular rotors and opens the door to the study of fundamental questions regarding criticality in systems with continuous degrees of freedom.
\end{abstract}

\maketitle

The ability to produce endofullerenes by molecular surgery\cite{komatsu2005encapsulation} has resulted in a number of exciting results, both experimental\cite{kurotobi2011single,krachmalnicoff2016dipolar} and theoretical.\cite{room2013infrared,xu2015general,felker2017accurate,felker2017electric,felker2017explaining,kalugina2017potential}
The generation of carbon nanotube ``peapods'' has also recently been shown to be possible.\cite{smith1998encapsulated,smith1999carbon,burteaux1999abundance,sloan2000size,monthioux2002filling,hornbaker2002mapping}
The combination of these ideas leads to endofullerene peapods: carbon nanotubes which contain fullerene cages with atoms or molecules trapped inside.\cite{hirahara2000one,sun2004synthesis,shiozawa2006filling,kitaura2007magnetism,nicholls2010direct,yang2010quantum,fritz2017nanoscale}

By treating these nanomolecular assemblies (NMAs) as fixed and rigid, we may study the motion of the atoms and molecules enclosed therein.
The resulting model is similar in some respects to that obtained from placing ultracold particles in an optical lattice,\cite{mazurenko2017cold,tai2017microscopy} and has previously been studied in that context.\cite{gorshkov2011tunable,abolins2011ground}
Nevertheless, there are some fundamental differences: in an NMA, the imprisoned entities may not move between sites as they do in an optical lattice, so there cannot be double occupation of a site; the spacing between adjacent sites is much smaller in an NMA (on the order of \SI{1}{nm})\cite{hornbaker2002mapping} than in a typical optical lattice (on the order of \SI{100}{nm});\cite{mazurenko2017cold} and the carbon walls of the fullerene cages shield the interactions between the captive particles.\cite{krachmalnicoff2016dipolar}

For an endofullerene peapod NMA model as described above, one is in principle left with translational, vibrational, and rotational degrees of freedom for the confined particles.
At very low temperatures, the translations and vibrations are restricted to their respective ground states, and only the rotational motion remains relevant, such that one can approximate the low-lying energy spectrum with an effective rotor Hamiltonian.
In the following, we therefore focus on the rotational degrees of freedom of molecules arranged in a linear chain.
Specifically, we choose dipolar linear rotor molecules (such as \ce{HF}, \ce{LiCl}, or \ce{CsI}) which interact pairwise through the (dimensionless) dipole--dipole potential
\begin{align}
	V_{ij}(\vec{e}_i, \vec{e}_j; \vec{r}_{ij})
	&= \vec{e}_i \cdot \vec{e}_j - 3 (\vec{e}_i \cdot \vec{r}_{ij}) (\vec{e}_j \cdot \vec{r}_{ij}),
		\label{eq:potential-full}
\end{align}
where $\vec{e}_i$ and $\vec{e}_j$ are unit vectors describing the orientation of two rotors, and $\vec{r}_{ij}$ is the unit vector in the direction from one rotor to the other.

In this communication, we propose a method for the calculation of ground state energies and wavefunctions for long one-dimensional systems of dipolar rotors using the density matrix renormalization group (DMRG).
Originally introduced by White in 1992,\cite{white1992density} the approach of DMRG has proven fruitful in a number of applications ranging from condensed matter physics\cite{schollwock2005density,schollwock2011density} to quantum chemistry.\cite{white1999ab,chan2011density,duperrouzel2015quantum}
Although it has been extended to the study of two-dimensional systems, finite temperature systems, and real-time evolution, DMRG excels at finding ground states of strongly-correlated one-dimensional systems.\cite{stoudenmire2012studying}

For small systems of this kind (up to around 10 rotors), sparse iterative methods for Hamiltonian diagonalization are sufficient to obtain a handful of low-lying eigenstates.
As the systems grow, the size of the many-body basis increases exponentially, and the problem quickly becomes intractable.
Hence, we turn to DMRG in order to grow the rotor chain under study to 50 rotors, which is made feasible by the matrix product state (MPS) wavefunction ansatz inherent to DMRG.
We accomplish this with the ITensor package, which allows us to efficiently formulate the Hamiltonian as a matrix product operator (MPO) and which contains an implementation of DMRG.\cite{itensor}
While existing publications have also examined many-body quantum systems with dipole--dipole interactions using DMRG,\cite{gorshkov2011tunable,ruhman2012nonlocal,manmana2013topological} due to their use of different geometries and focus on mapping to other model systems, they do not capture the full physics of interacting molecules under quantum rotation.

For $N$ identical rotors with rotational constant $B$ and dipole moment $\mu$, the general Hamiltonian is
\begin{align}
	\hat{H}
	&= \frac{B}{\hbar^2} \sum_{i=1}^N \hat{\ell}_i^2 + \frac{\mu^2}{4 \pi \epsilon_0} \sum_{i=2}^N \sum_{j=1}^{i-1} \frac{\hat{V}_{ij}}{r_{ij}^3},
\end{align}
where $r_{ij}$ is the distance between rotors $i$ and $j$.
Since a peapod NMA is inherently linear, without loss of generality, we may place the rotors along the $z$ axis and express the potential operator compactly as
\begin{align}
	\hat{V}_{ij}^{(z)}
	&= \hat{x}_i \hat{x}_j + \hat{y}_i \hat{y}_j - 2 \hat{z}_i \hat{z}_j.
\end{align}
Because of the regular structure of a peapod NMA, we space the rotors evenly and write the Hamiltonian as
\begin{align}
	\frac{\hat{H}}{B}
	&= \sum_{i=1}^N \frac{\hat{\ell}_i^2}{\hbar^2} + \frac{1}{R^3} \sum_{i=2}^N \sum_{j=1}^{i-1} \frac{\hat{V}_{ij}^{(z)}}{(i-j)^3},
		\label{eq:hamiltonian-z}
\end{align}
where
\begin{align}
	R
	&= r \left( \frac{4 \pi \epsilon_0 B}{\mu^2} \right)^\frac{1}{3},
\end{align}
$r$ is the distance between adjacent rotors (the lattice spacing), and we have taken this opportunity to non-dimensionalize the Hamiltonian.
Since all the physical properties appear only in $R$, the one-parameter form of the Hamiltonian allows us to explore the entire realm of physical realizations of this model by scanning a single parameter.
A smaller value of $R$ results in stronger interactions, potentially caused by a larger dipole moment, a smaller rotational constant, or a smaller inter-particle separation.

A natural one-body basis for this problem is that of the spherical harmonics $\ket{\ell_i \, m_i}$, in which the squared angular momentum operator $\hat{\ell}_i^2$ is diagonal:
\begin{align}
	\hat{\ell}_i^2 \ket{\ell_i \, m_i}
	&= \hbar^2 \ell_i (\ell_i + 1) \ket{\ell_i \, m_i}.
\end{align}
Although in principle this basis is infinite, in order to carry out any calculations, it will need to be truncated at a finite $\lmax$ so that it is large enough to accurately represent the quantities in question, but no larger.

Thanks to the form of the potential operator, the Hamiltonian conserves
\begin{subequations}
\begin{align}
	\qnlp
	&\equiv \sum_{i=1}^N \ell_i \pmod{2} \\
\intertext{and}
	\qnm
	&= \sum_{i=1}^N m_i.
\end{align}
\end{subequations}
In order to exploit the block-diagonal structure of the Hamiltonian in DMRG, we must make explicit use of these good quantum numbers.
That is, we need to express the potential operator in terms of one-body operators that only change the quantum numbers $\qnlp$ and $\qnm$ by a definite amount, termed the ``flux''.
This makes it possible to construct both the wavefunction MPS and Hamiltonian MPO as sparse objects, reducing the amount of storage required and significantly accelerating the calculation.\cite{schollwock2011density}
Terms like $\hat{x}_i \hat{x}_j$ do not suffice, because the position operators ($\hat{x}_i$, $\hat{y}_i$, $\hat{z}_i$) do not have a well-defined flux.
The action of one of these operators on a state with definite $\qnlp$ and $\qnm$ quantum numbers does not result in a state with definite $\qnlp'$ and $\qnm'$ values.

The ladder operators $\hat{\ell}_i^\pm$ and $\hat{m}_i^\pm$, which raise and lower $\ell_i$ and $m_i$, are obvious candidates for building blocks, as their flux is immediately evident.
The latter operators have the well-known form
\begin{align}
	\hat{m}_i^\pm
	&= \hat{\ell}_{i,x} \pm i \hat{\ell}_{i,y}
\end{align}
and they act as\cite{ballentine1990quantum}
\begin{align}
	& \hat{m}_i^\pm \ket{\ell_i \, m_i} \notag \\
	&= \hbar \sqrt{(\ell_i \pm m_i + 1) (\ell_i \mp m_i)} \ket{\ell_i , m_i \pm 1}.
\end{align}
On the other hand, the ladder operators for $\ell_i$ do not appear to have been as deeply analyzed.
There exist the definitions\cite{liu2010raising}
\begin{subequations}
\begin{align}
	\hbar \hat{R}_{i,z}
	&= i (\hat{x}_i \hat{\ell}_{i,y} - \hat{y}_i \hat{\ell}_{i,x})
		+ \frac{\hat{z}_i}{2} \left( \hbar + \sqrt{4 \hat{\ell}_i^2 + \hbar^2} \right) \\
\intertext{and}
	\hbar \hat{Q}_{i,z}
	&= i (\hat{x}_i \hat{\ell}_{i,y} - \hat{y}_i \hat{\ell}_{i,x})
		+ \frac{\hat{z}_i}{2} \left( \hbar - \sqrt{4 \hat{\ell}_i^2 + \hbar^2} \right),
\end{align}
\end{subequations}
but unfortunately $\hat{R}_{i,z}\adj \ne \hat{Q}_{i,z}$.
We instead introduce the operators
\begin{align}
	\hat{\ell}_i^\pm
	&= \frac{\hbar}{2} \hat{z}_i \left( 1 \pm \frac{\hbar}{\sqrt{4 \hat{\ell}_i^2 + \hbar^2}} \right) \notag \\
	&\qquad
		\pm i (\hat{x}_i \hat{\ell}_{i,y} - \hat{y}_i \hat{\ell}_{i,x}) \frac{\hbar}{\sqrt{4 \hat{\ell}_i^2 + \hbar^2}},
\end{align}
which are intimately related to $\hat{R}_{i,z}$ and $\hat{Q}_{i,z}$, but satisfy $(\hat{\ell}_i^+)\adj = \hat{\ell}_i^-$.

From this definition, it follows that
\begin{subequations}
\begin{align}
	\hat{x}_i
	&= \frac{1}{2 \hbar^2} \comm{(\hat{\ell}_i^+ + \hat{\ell}_i^-)}{(\hat{m}_i^+ - \hat{m}_i^-)}, \\
	\hat{y}_i
	&= \frac{1}{2i \hbar^2} \comm{(\hat{\ell}_i^+ + \hat{\ell}_i^-)}{(\hat{m}_i^+ + \hat{m}_i^-)}, \\
\intertext{and}
	\hat{z}_i
	&= \frac{1}{\hbar} (\hat{\ell}_i^+ + \hat{\ell}_i^-).
\end{align}
\end{subequations}
The clean and concise form of these expressions suggests that our choice of the ladder operators $\hat{\ell}_i^\pm$ is an appropriate one.
The potential from Eq. \ref{eq:potential-full} may then be written as
\begin{align}
	\hat{V}_{ij}
	&= (1 - 3 r_{ij,z}^2) \hat{B}_i^0 \hat{B}_j^0 \notag \\
	&\qquad
		- \frac{1}{4} \left[
			(1 - 3 r_{ij,z}^2) \hat{B}_i^- \hat{B}_j^+
			+ 3 r_{ij,\perp}^2 \hat{B}_i^- \hat{B}_j^-
		\right. \notag \\
	&\qquad\qquad\quad
		\left.
			+ 6 r_{ij,\perp} r_{ij,z} (\hat{B}_i^- \hat{B}_j^0 + \hat{B}_i^0 \hat{B}_j^-)
		\right. \notag \\
	&\qquad\qquad\quad
		\left.
			+ \mathrm{h.c.}
		\right],
\end{align}
where
\begin{subequations}
\begin{align}
	\hat{B}_i^\pm
	&= \pm\frac{1}{\hbar^2} \left( \comm{\hat{\ell}_i^+}{\hat{m}_i^\pm} + \comm{\hat{\ell}_i^-}{\hat{m}_i^\pm} \right), \\
	\hat{B}_i^0
	&= \frac{1}{\hbar} (\hat{\ell}_i^+ + \hat{\ell}_i^-),
\end{align}
\end{subequations}
and
\begin{align}
	r_{ij,\perp}
	&= r_{ij,x} + i r_{ij,y}.
\end{align}
The simplified form for rotors aligned along the $z$ axis is
\begin{align}
	\hat{V}_{ij}^{(z)}
	&= -2 \hat{B}_i^0 \hat{B}_j^0
		+ \frac{1}{2} \left[
			\hat{B}_i^- \hat{B}_j^+
			+ \mathrm{h.c.}
		\right].
\end{align}
When written in this form, the potential operator may be constructed as a sparse MPO.

\begin{figure}[t]
	\includegraphics{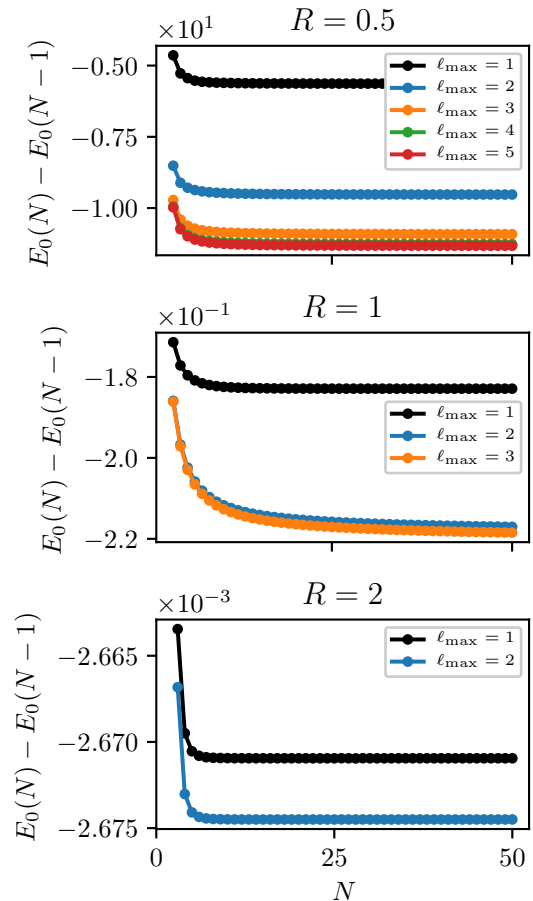}
	\caption{
		Chemical potential of rotor chains of length $N$.
		Several curves with different $\lmax$ are shown to demonstrate the effect of basis truncation.
		The chemical potential at $R = 1$ takes longer than the others to plateau.
	}
	\label{fig:chemical_potential}
\end{figure}

The primary result of the DMRG routine is the ground state energy $E_0$.
As these energies are expected to decrease with increasing system size $N$, we present them in the form of chemical potentials in Fig. \ref{fig:chemical_potential}.
It is evident that a smaller value of $R$ (stronger interactions) requires a larger $\lmax$ (more basis states), as expected.
For sufficiently large systems, we expect on physical grounds that the addition of a single particle will result in a constant decrease in the energy of the system, regardless of the system size.
In other words, because a newly added rotor should only be substantially correlated with finitely many rotors on the end of the system, the chemical potential should tend to a constant in the large $N$ limit.
For $R = 0.5$ and $R = 2$, this limit is reached by 25 rotors, but for $R = 1$, the chemical potential continues to change even at 50 rotors, indicating longer-ranged correlations.

\begin{figure}[t]
	\includegraphics{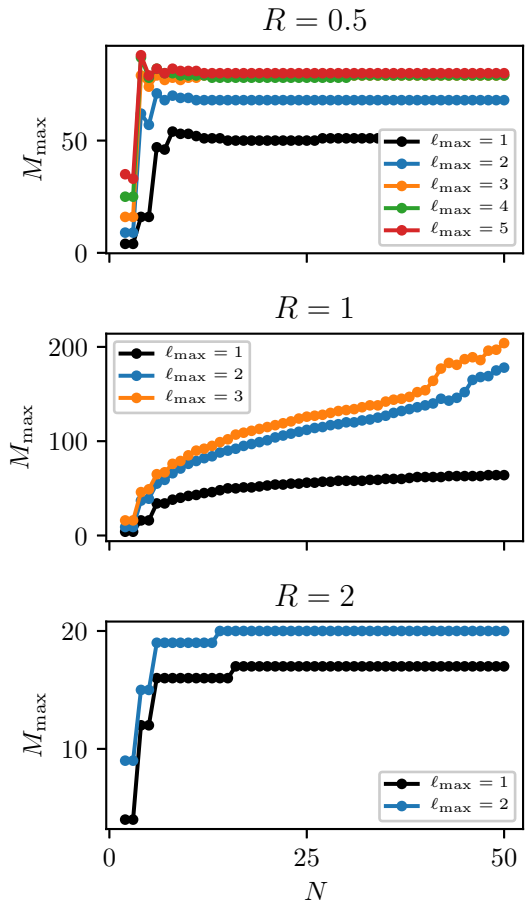}
	\caption{
		Maximum MPS bond dimension for rotor chains of length $N$.
		Several curves with different $\lmax$ are shown to demonstrate the effect of basis truncation.
		The bond dimension at $R = 1$ is larger than the others and not constant by $N = 50$.
	}
	\label{fig:bond}
\end{figure}

\begin{figure}[t]
	\includegraphics{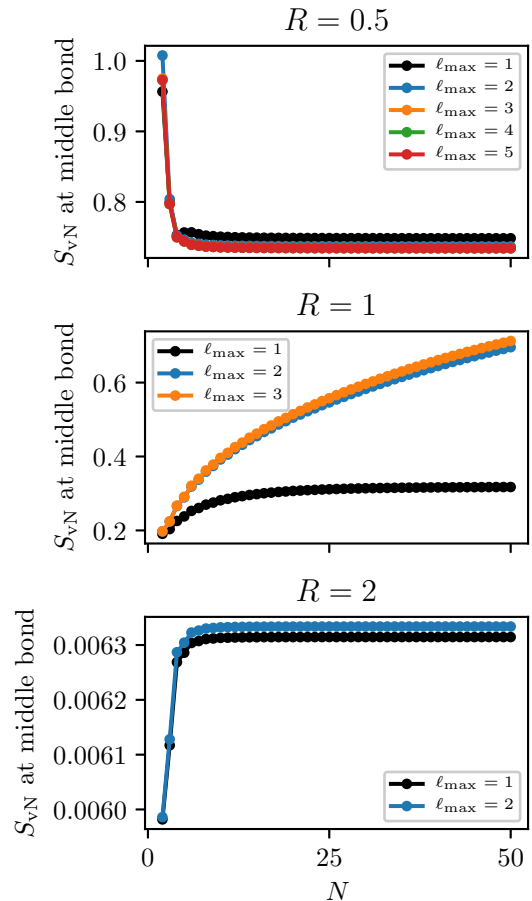}
	\caption{
		Von Neumann entanglement entropy of rotor chains of length $N$.
		Several curves with different $\lmax$ are shown to demonstrate the effect of basis truncation.
		The entropy at $R = 1$ is still increasing by $N = 50$.
	}
	\label{fig:entropy}
\end{figure}

The MPS representation of the wavefunction has a ``bond dimension'' $M$ at the bond between any two adjacent rotors, and this number bounds the possible amount of entanglement between the rotors on either side of the bond.
The maximum bond dimension $M_\text{max}$, shown in Fig. \ref{fig:bond}, is the largest bond dimension across the entire MPS, and it is indicative of the amount of long-range correlations in the state.
That $M_\text{max}$ plateaus quickly for $R = 0.5$ and $R = 2$ implies the presence of only short-range correlations, but the same cannot be said for $R = 1$.
The von Neumann entanglement entropy $S_\text{vN}$ for the partitioning of the system into halves behaves similarly to the bond dimension, as shown in Fig. \ref{fig:entropy}.

\begin{figure}[t]
	\includegraphics{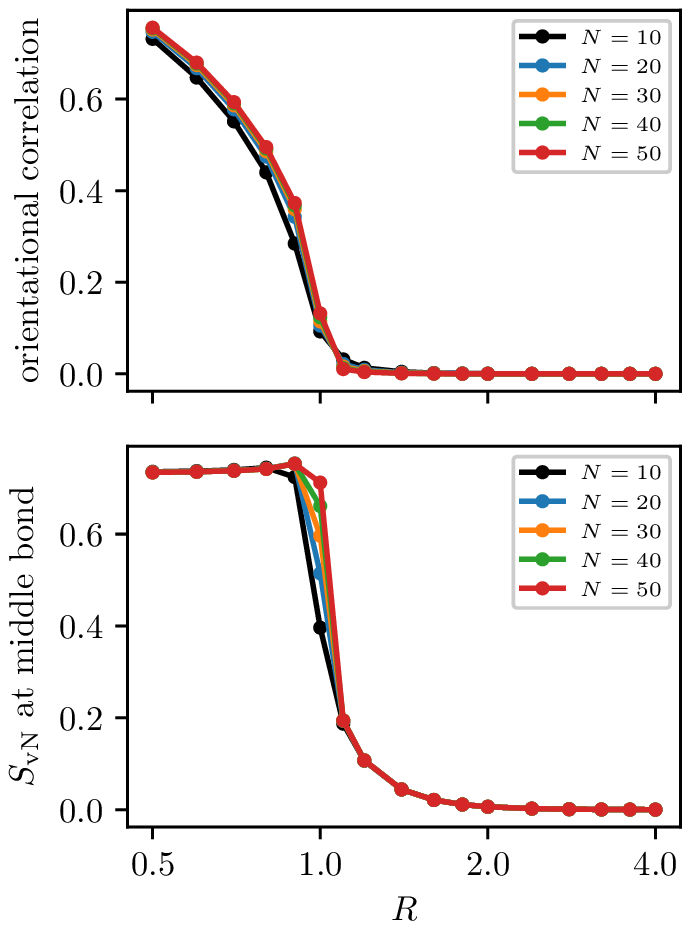}
	\caption{
		Comparison of the orientational correlation (top panel) and the entanglement entropy (bottom panel) for rotor chains of size $N$ across the apparent phase transition.
	}
	\label{fig:phase}
\end{figure}

The quantum rotor model, which resembles the model used in the present communication, but lacks the anisotropic term in Eq. \ref{eq:potential-full}, is known to have no ordered phase in one dimension and therefore no phase transition.\cite{sachdev2011quantum}
In light of this, the observed anomalies at $R = 1$ are peculiar, but it has been suggested that the breaking of rotational symmetry in the anisotropic model is responsible for a second-order phase transition between ordered and disordered phases.\cite{abolins2011ground}
This is corroborated by the sudden change in both the expectation value of the orientational correlation operator,
\begin{align}
	\frac{2}{N (N-1)} \sum_{i=2}^N \sum_{j=1}^{i-1} \hat{\vec{e}}_i \cdot \hat{\vec{e}}_j,
\end{align}
and the von Neumann entanglement entropy $S_\text{vN}$ near $R = 1$, as demonstrated in Fig. \ref{fig:phase}.
Of the two, it seems that the latter is a sharper indicator of the apparent phase transition.

At a second order phase transition, the spatial correlation length should diverge.\cite{cardy2002scaling}
Thus, in the near future we plan to examine the behaviour of the correlation length around $R = 1$ to confirm the existence of the transition and identify the value of the critical parameter $R_\text{c}$.
We then hope to extract the central charge of the relevant conformal field theory for the critical system.\cite{calabrese2009entanglement}

Because the Hamiltonian in Eq. \ref{eq:hamiltonian-z} is block-diagonal, we can also use DMRG to target the ground state of any symmetry block.
The $\qnlp = 0, \qnm = 0$ block contains the ground state of the entire Hamiltonian, but the $\qnlp = 1, \qnm = 0$ and $\qnlp = 1, \qnm = \pm 1$ blocks are also of interest, because their states $\ket{n_{\qnlp,\qnm}}$ are reachable from the ground state $\ket{0_{\qnlp = 0, \qnm = 0}}$ by application of the $\hat{x}_i$ and $\hat{z}_i$ operators.
That is, while most transition dipole moments (TDMs) are forbidden by symmetry, the moments
\begin{subequations}
\begin{align}
	| \mel{n_{\qnlp = 1, \qnm = 0}}{\frac{1}{N} \sum_{i=1}^N \hat{z}_i}{0_{\qnlp = 0, \qnm = 0}} |^2 \\
\intertext{and}
	| \mel{n_{\qnlp = 1, \qnm = \pm 1}}{\frac{1}{N} \sum_{i=1}^N \hat{x}_i}{0_{\qnlp = 0, \qnm = 0}} |^2.
\end{align}
\end{subequations}
do not necessarily vanish.

Computed energy differences $\Delta E_0$ and TDMs for $R = 2$ are listed in Table \ref{tab:transitions}.
Despite this calculation providing only two peaks of a dipole excitation spectrum for each system size, it lays the foundation for a series of more involved calculations which can reveal more information from the spectrum.
We propose two complementary approaches for this.
The first involves the direct calculation of excited states within the identified symmetry blocks.
From these, more energy differences and TDMs can be computed, gradually populating a stick spectrum.
The second requires time evolution of the ground state to obtain a correlation function, followed by a Fourier transform which yields a spectrum with finite resolution.
Both approaches are presently possible using standard extensions to DMRG for excited states and real-time evolution.\cite{schollwock2011density}

\begin{table}[t]
\caption{
		Energy differences and transition dipole moments for rotor chains of length $N$ at $R = 2$.
		Each value is computed between the ground state of the entire Hamiltonian and the ground state of the specified block at $\lmax = 12$ and with 30 DMRG sweeps.
	}
	\begin{tabular}{c c r c c}
		\hline
		\hline
		$N$ & $\qnlp$ & $\qnm$ & $\Delta E_0$ & TDM \\
		\hline
		\multirow{2}{*}{$5$} & $1$ & $0$ & $1.856$ & $5.935 \times 10^{-2}$ \\
		& $1$ & $\pm 1$ & $1.950$ & $3.117 \times 10^{-4}$ \\
		\multirow{2}{*}{$10$} & $1$ & $0$ & $1.843$ & $2.296 \times 10^{-2}$ \\
		& $1$ & $\pm 1$ & $1.950$ & $7.340 \times 10^{-5}$ \\
		\multirow{2}{*}{$15$} & $1$ & $0$ & $1.824$ & $1.611 \times 10^{-2}$ \\
		& $1$ & $\pm 1$ & $1.948$ & $3.504 \times 10^{-5}$ \\
		\hline
		\hline
	\end{tabular}
		\label{tab:transitions}
\end{table}

An important extension to the above model is the addition of translational motion for the rotor molecules.\cite{felker2017accurate}
The primary difficulty in implementing this change is the dynamical nature of the intermolecular separations: the classical parameter $R$ must be augmented by quantum mechanical operators which describe the deviations from the cage centers.
Although it is not currently clear how such an implementation would look, it is likely to involve a model for the rotation--translation coupling which can be expressed in terms of raising and lowering operators for the site-local translational states.

This research was supported by the Natural Sciences and Engineering Research Council of Canada (NSERC), the Ontario Ministry of Research and Innovation (MRI), the Canada Research Chair program, the Canada Foundation for Innovation (CFI), and the Canada First Research Excellence Fund (CFREF).
We thank Martin Ganahl and Roger Melko for  informative discussions.

\bibliography{paper}

\begin{thebibliography}{41}%
\makeatletter
\providecommand \@ifxundefined [1]{%
 \@ifx{#1\undefined}
}%
\providecommand \@ifnum [1]{%
 \ifnum #1\expandafter \@firstoftwo
 \else \expandafter \@secondoftwo
 \fi
}%
\providecommand \@ifx [1]{%
 \ifx #1\expandafter \@firstoftwo
 \else \expandafter \@secondoftwo
 \fi
}%
\providecommand \natexlab [1]{#1}%
\providecommand \enquote  [1]{``#1''}%
\providecommand \bibnamefont  [1]{#1}%
\providecommand \bibfnamefont [1]{#1}%
\providecommand \citenamefont [1]{#1}%
\providecommand \href@noop [0]{\@secondoftwo}%
\providecommand \href [0]{\begingroup \@sanitize@url \@href}%
\providecommand \@href[1]{\@@startlink{#1}\@@href}%
\providecommand \@@href[1]{\endgroup#1\@@endlink}%
\providecommand \@sanitize@url [0]{\catcode `\\12\catcode `\$12\catcode
  `\&12\catcode `\#12\catcode `\^12\catcode `\_12\catcode `\%12\relax}%
\providecommand \@@startlink[1]{}%
\providecommand \@@endlink[0]{}%
\providecommand \url  [0]{\begingroup\@sanitize@url \@url }%
\providecommand \@url [1]{\endgroup\@href {#1}{\urlprefix }}%
\providecommand \urlprefix  [0]{URL }%
\providecommand \Eprint [0]{\href }%
\providecommand \doibase [0]{http://dx.doi.org/}%
\providecommand \selectlanguage [0]{\@gobble}%
\providecommand \bibinfo  [0]{\@secondoftwo}%
\providecommand \bibfield  [0]{\@secondoftwo}%
\providecommand \translation [1]{[#1]}%
\providecommand \BibitemOpen [0]{}%
\providecommand \bibitemStop [0]{}%
\providecommand \bibitemNoStop [0]{.\EOS\space}%
\providecommand \EOS [0]{\spacefactor3000\relax}%
\providecommand \BibitemShut  [1]{\csname bibitem#1\endcsname}%
\let\auto@bib@innerbib\@empty
\bibitem [{\citenamefont {Komatsu}, \citenamefont {Murata},\ and\ \citenamefont
  {Murata}(2005)}]{komatsu2005encapsulation}%
  \BibitemOpen
  \bibfield  {author} {\bibinfo {author} {\bibfnamefont {K.}~\bibnamefont
  {Komatsu}}, \bibinfo {author} {\bibfnamefont {M.}~\bibnamefont {Murata}}, \
  and\ \bibinfo {author} {\bibfnamefont {Y.}~\bibnamefont {Murata}},\
  }\href@noop {} {\bibfield  {journal} {\bibinfo  {journal} {Science}\ }\textbf
  {\bibinfo {volume} {307}},\ \bibinfo {pages} {238} (\bibinfo {year}
  {2005})}\BibitemShut {NoStop}%
\bibitem [{\citenamefont {Kurotobi}\ and\ \citenamefont
  {Murata}(2011)}]{kurotobi2011single}%
  \BibitemOpen
  \bibfield  {author} {\bibinfo {author} {\bibfnamefont {K.}~\bibnamefont
  {Kurotobi}}\ and\ \bibinfo {author} {\bibfnamefont {Y.}~\bibnamefont
  {Murata}},\ }\href@noop {} {\bibfield  {journal} {\bibinfo  {journal}
  {Science}\ }\textbf {\bibinfo {volume} {333}},\ \bibinfo {pages} {613}
  (\bibinfo {year} {2011})}\BibitemShut {NoStop}%
\bibitem [{\citenamefont {Krachmalnicoff}\ \emph {et~al.}(2016)\citenamefont
  {Krachmalnicoff}, \citenamefont {Bounds}, \citenamefont {Mamone},
  \citenamefont {Alom}, \citenamefont {Concistr{\`e}}, \citenamefont {Meier},
  \citenamefont {Kou{\v{r}}il}, \citenamefont {Light}, \citenamefont {Johnson},
  \citenamefont {Rols} \emph {et~al.}}]{krachmalnicoff2016dipolar}%
  \BibitemOpen
  \bibfield  {author} {\bibinfo {author} {\bibfnamefont {A.}~\bibnamefont
  {Krachmalnicoff}}, \bibinfo {author} {\bibfnamefont {R.}~\bibnamefont
  {Bounds}}, \bibinfo {author} {\bibfnamefont {S.}~\bibnamefont {Mamone}},
  \bibinfo {author} {\bibfnamefont {S.}~\bibnamefont {Alom}}, \bibinfo {author}
  {\bibfnamefont {M.}~\bibnamefont {Concistr{\`e}}}, \bibinfo {author}
  {\bibfnamefont {B.}~\bibnamefont {Meier}}, \bibinfo {author} {\bibfnamefont
  {K.}~\bibnamefont {Kou{\v{r}}il}}, \bibinfo {author} {\bibfnamefont {M.~E.}\
  \bibnamefont {Light}}, \bibinfo {author} {\bibfnamefont {M.~R.}\ \bibnamefont
  {Johnson}}, \bibinfo {author} {\bibfnamefont {S.}~\bibnamefont {Rols}},
  \emph {et~al.},\ }\href@noop {} {\bibfield  {journal} {\bibinfo  {journal}
  {Nat. Chem.}\ }\textbf {\bibinfo {volume} {8}},\ \bibinfo {pages} {953}
  (\bibinfo {year} {2016})}\BibitemShut {NoStop}%
\bibitem [{\citenamefont {R{\~o}{\~o}m}\ \emph {et~al.}(2013)\citenamefont
  {R{\~o}{\~o}m}, \citenamefont {Peedu}, \citenamefont {Ge}, \citenamefont
  {H{\"u}vonen}, \citenamefont {Nagel}, \citenamefont {Ye}, \citenamefont {Xu},
  \citenamefont {Ba{\v{c}}i{\'c}}, \citenamefont {Mamone}, \citenamefont
  {Levitt} \emph {et~al.}}]{room2013infrared}%
  \BibitemOpen
  \bibfield  {author} {\bibinfo {author} {\bibfnamefont {T.}~\bibnamefont
  {R{\~o}{\~o}m}}, \bibinfo {author} {\bibfnamefont {L.}~\bibnamefont {Peedu}},
  \bibinfo {author} {\bibfnamefont {M.}~\bibnamefont {Ge}}, \bibinfo {author}
  {\bibfnamefont {D.}~\bibnamefont {H{\"u}vonen}}, \bibinfo {author}
  {\bibfnamefont {U.}~\bibnamefont {Nagel}}, \bibinfo {author} {\bibfnamefont
  {S.}~\bibnamefont {Ye}}, \bibinfo {author} {\bibfnamefont {M.}~\bibnamefont
  {Xu}}, \bibinfo {author} {\bibfnamefont {Z.}~\bibnamefont {Ba{\v{c}}i{\'c}}},
  \bibinfo {author} {\bibfnamefont {S.}~\bibnamefont {Mamone}}, \bibinfo
  {author} {\bibfnamefont {M.~H.}\ \bibnamefont {Levitt}},  \emph {et~al.},\
  }\href@noop {} {\bibfield  {journal} {\bibinfo  {journal} {Philos. Trans. R.
  Soc., A}\ }\textbf {\bibinfo {volume} {371}},\ \bibinfo {pages} {20110631}
  (\bibinfo {year} {2013})}\BibitemShut {NoStop}%
\bibitem [{\citenamefont {Xu}, \citenamefont {Ye},\ and\ \citenamefont
  {Bačić}(2015)}]{xu2015general}%
  \BibitemOpen
  \bibfield  {author} {\bibinfo {author} {\bibfnamefont {M.}~\bibnamefont
  {Xu}}, \bibinfo {author} {\bibfnamefont {S.}~\bibnamefont {Ye}}, \ and\
  \bibinfo {author} {\bibfnamefont {Z.}~\bibnamefont {Bačić}},\ }\href@noop
  {} {\bibfield  {journal} {\bibinfo  {journal} {J. Phys. Chem. Lett.}\
  }\textbf {\bibinfo {volume} {6}},\ \bibinfo {pages} {3721} (\bibinfo {year}
  {2015})}\BibitemShut {NoStop}%
\bibitem [{\citenamefont {Felker}\ and\ \citenamefont
  {Ba{\v{c}}i{\'c}}(2017{\natexlab{a}})}]{felker2017accurate}%
  \BibitemOpen
  \bibfield  {author} {\bibinfo {author} {\bibfnamefont {P.~M.}\ \bibnamefont
  {Felker}}\ and\ \bibinfo {author} {\bibfnamefont {Z.}~\bibnamefont
  {Ba{\v{c}}i{\'c}}},\ }\href@noop {} {\bibfield  {journal} {\bibinfo
  {journal} {Chem. Phys. Lett.}\ }\textbf {\bibinfo {volume} {683}},\ \bibinfo
  {pages} {172} (\bibinfo {year} {2017}{\natexlab{a}})}\BibitemShut {NoStop}%
\bibitem [{\citenamefont {Felker}\ and\ \citenamefont
  {Ba{\v{c}}i{\'c}}(2017{\natexlab{b}})}]{felker2017electric}%
  \BibitemOpen
  \bibfield  {author} {\bibinfo {author} {\bibfnamefont {P.~M.}\ \bibnamefont
  {Felker}}\ and\ \bibinfo {author} {\bibfnamefont {Z.}~\bibnamefont
  {Ba{\v{c}}i{\'c}}},\ }\href@noop {} {\bibfield  {journal} {\bibinfo
  {journal} {J. Chem. Phys.}\ }\textbf {\bibinfo {volume} {146}},\ \bibinfo
  {pages} {084303} (\bibinfo {year} {2017}{\natexlab{b}})}\BibitemShut
  {NoStop}%
\bibitem [{\citenamefont {Felker}\ \emph {et~al.}(2017)\citenamefont {Felker},
  \citenamefont {Vl{\v{c}}ek}, \citenamefont {Hietanen}, \citenamefont
  {FitzGerald}, \citenamefont {Neuhauser},\ and\ \citenamefont
  {Ba{\v{c}}i{\'c}}}]{felker2017explaining}%
  \BibitemOpen
  \bibfield  {author} {\bibinfo {author} {\bibfnamefont {P.~M.}\ \bibnamefont
  {Felker}}, \bibinfo {author} {\bibfnamefont {V.}~\bibnamefont {Vl{\v{c}}ek}},
  \bibinfo {author} {\bibfnamefont {I.}~\bibnamefont {Hietanen}}, \bibinfo
  {author} {\bibfnamefont {S.}~\bibnamefont {FitzGerald}}, \bibinfo {author}
  {\bibfnamefont {D.}~\bibnamefont {Neuhauser}}, \ and\ \bibinfo {author}
  {\bibfnamefont {Z.}~\bibnamefont {Ba{\v{c}}i{\'c}}},\ }\href@noop {}
  {\bibfield  {journal} {\bibinfo  {journal} {Phys. Chem. Chem. Phys.}\
  }\textbf {\bibinfo {volume} {19}},\ \bibinfo {pages} {31274} (\bibinfo {year}
  {2017})}\BibitemShut {NoStop}%
\bibitem [{\citenamefont {Kalugina}\ and\ \citenamefont
  {Roy}(2017)}]{kalugina2017potential}%
  \BibitemOpen
  \bibfield  {author} {\bibinfo {author} {\bibfnamefont {Y.~N.}\ \bibnamefont
  {Kalugina}}\ and\ \bibinfo {author} {\bibfnamefont {P.-N.}\ \bibnamefont
  {Roy}},\ }\href@noop {} {\bibfield  {journal} {\bibinfo  {journal} {J. Chem.
  Phys.}\ }\textbf {\bibinfo {volume} {147}},\ \bibinfo {pages} {244303}
  (\bibinfo {year} {2017})}\BibitemShut {NoStop}%
\bibitem [{\citenamefont {Smith}, \citenamefont {Monthioux},\ and\
  \citenamefont {Luzzi}(1998)}]{smith1998encapsulated}%
  \BibitemOpen
  \bibfield  {author} {\bibinfo {author} {\bibfnamefont {B.~W.}\ \bibnamefont
  {Smith}}, \bibinfo {author} {\bibfnamefont {M.}~\bibnamefont {Monthioux}}, \
  and\ \bibinfo {author} {\bibfnamefont {D.~E.}\ \bibnamefont {Luzzi}},\
  }\href@noop {} {\bibfield  {journal} {\bibinfo  {journal} {Nature}\ }\textbf
  {\bibinfo {volume} {396}},\ \bibinfo {pages} {323} (\bibinfo {year}
  {1998})}\BibitemShut {NoStop}%
\bibitem [{\citenamefont {Smith}, \citenamefont {Monthioux},\ and\
  \citenamefont {Luzzi}(1999)}]{smith1999carbon}%
  \BibitemOpen
  \bibfield  {author} {\bibinfo {author} {\bibfnamefont {B.~W.}\ \bibnamefont
  {Smith}}, \bibinfo {author} {\bibfnamefont {M.}~\bibnamefont {Monthioux}}, \
  and\ \bibinfo {author} {\bibfnamefont {D.~E.}\ \bibnamefont {Luzzi}},\
  }\href@noop {} {\bibfield  {journal} {\bibinfo  {journal} {Chem. Phys.
  Lett.}\ }\textbf {\bibinfo {volume} {315}},\ \bibinfo {pages} {31} (\bibinfo
  {year} {1999})}\BibitemShut {NoStop}%
\bibitem [{\citenamefont {Burteaux}\ \emph {et~al.}(1999)\citenamefont
  {Burteaux}, \citenamefont {Claye}, \citenamefont {Smith}, \citenamefont
  {Monthioux}, \citenamefont {Luzzi},\ and\ \citenamefont
  {Fischer}}]{burteaux1999abundance}%
  \BibitemOpen
  \bibfield  {author} {\bibinfo {author} {\bibfnamefont {B.}~\bibnamefont
  {Burteaux}}, \bibinfo {author} {\bibfnamefont {A.}~\bibnamefont {Claye}},
  \bibinfo {author} {\bibfnamefont {B.~W.}\ \bibnamefont {Smith}}, \bibinfo
  {author} {\bibfnamefont {M.}~\bibnamefont {Monthioux}}, \bibinfo {author}
  {\bibfnamefont {D.~E.}\ \bibnamefont {Luzzi}}, \ and\ \bibinfo {author}
  {\bibfnamefont {J.~E.}\ \bibnamefont {Fischer}},\ }\href@noop {} {\bibfield
  {journal} {\bibinfo  {journal} {Chem. Phys. Lett.}\ }\textbf {\bibinfo
  {volume} {310}},\ \bibinfo {pages} {21} (\bibinfo {year} {1999})}\BibitemShut
  {NoStop}%
\bibitem [{\citenamefont {Sloan}\ \emph {et~al.}(2000)\citenamefont {Sloan},
  \citenamefont {Dunin-Borkowski}, \citenamefont {Hutchison}, \citenamefont
  {Coleman}, \citenamefont {Williams}, \citenamefont {Claridge}, \citenamefont
  {York}, \citenamefont {Xu}, \citenamefont {Bailey}, \citenamefont {Brown}
  \emph {et~al.}}]{sloan2000size}%
  \BibitemOpen
  \bibfield  {author} {\bibinfo {author} {\bibfnamefont {J.}~\bibnamefont
  {Sloan}}, \bibinfo {author} {\bibfnamefont {R.~E.}\ \bibnamefont
  {Dunin-Borkowski}}, \bibinfo {author} {\bibfnamefont {J.~L.}\ \bibnamefont
  {Hutchison}}, \bibinfo {author} {\bibfnamefont {K.~S.}\ \bibnamefont
  {Coleman}}, \bibinfo {author} {\bibfnamefont {V.~C.}\ \bibnamefont
  {Williams}}, \bibinfo {author} {\bibfnamefont {J.~B.}\ \bibnamefont
  {Claridge}}, \bibinfo {author} {\bibfnamefont {A.~P.~E.}\ \bibnamefont
  {York}}, \bibinfo {author} {\bibfnamefont {C.}~\bibnamefont {Xu}}, \bibinfo
  {author} {\bibfnamefont {S.~R.}\ \bibnamefont {Bailey}}, \bibinfo {author}
  {\bibfnamefont {G.}~\bibnamefont {Brown}},  \emph {et~al.},\ }\href@noop {}
  {\bibfield  {journal} {\bibinfo  {journal} {Chem. Phys. Lett.}\ }\textbf
  {\bibinfo {volume} {316}},\ \bibinfo {pages} {191} (\bibinfo {year}
  {2000})}\BibitemShut {NoStop}%
\bibitem [{\citenamefont {Monthioux}(2002)}]{monthioux2002filling}%
  \BibitemOpen
  \bibfield  {author} {\bibinfo {author} {\bibfnamefont {M.}~\bibnamefont
  {Monthioux}},\ }\href@noop {} {\bibfield  {journal} {\bibinfo  {journal}
  {Carbon}\ }\textbf {\bibinfo {volume} {40}},\ \bibinfo {pages} {1809}
  (\bibinfo {year} {2002})}\BibitemShut {NoStop}%
\bibitem [{\citenamefont {Hornbaker}\ \emph {et~al.}(2002)\citenamefont
  {Hornbaker}, \citenamefont {Kahng}, \citenamefont {Misra}, \citenamefont
  {Smith}, \citenamefont {Johnson}, \citenamefont {Mele}, \citenamefont
  {Luzzi},\ and\ \citenamefont {Yazdani}}]{hornbaker2002mapping}%
  \BibitemOpen
  \bibfield  {author} {\bibinfo {author} {\bibfnamefont {D.~J.}\ \bibnamefont
  {Hornbaker}}, \bibinfo {author} {\bibfnamefont {S.-J.}\ \bibnamefont
  {Kahng}}, \bibinfo {author} {\bibfnamefont {S.}~\bibnamefont {Misra}},
  \bibinfo {author} {\bibfnamefont {B.~W.}\ \bibnamefont {Smith}}, \bibinfo
  {author} {\bibfnamefont {A.~T.}\ \bibnamefont {Johnson}}, \bibinfo {author}
  {\bibfnamefont {E.~J.}\ \bibnamefont {Mele}}, \bibinfo {author}
  {\bibfnamefont {D.~E.}\ \bibnamefont {Luzzi}}, \ and\ \bibinfo {author}
  {\bibfnamefont {A.}~\bibnamefont {Yazdani}},\ }\href@noop {} {\bibfield
  {journal} {\bibinfo  {journal} {Science}\ }\textbf {\bibinfo {volume}
  {295}},\ \bibinfo {pages} {828} (\bibinfo {year} {2002})}\BibitemShut
  {NoStop}%
\bibitem [{\citenamefont {Hirahara}\ \emph {et~al.}(2000)\citenamefont
  {Hirahara}, \citenamefont {Suenaga}, \citenamefont {Bandow}, \citenamefont
  {Kato}, \citenamefont {Okazaki}, \citenamefont {Shinohara},\ and\
  \citenamefont {Iijima}}]{hirahara2000one}%
  \BibitemOpen
  \bibfield  {author} {\bibinfo {author} {\bibfnamefont {K.}~\bibnamefont
  {Hirahara}}, \bibinfo {author} {\bibfnamefont {K.}~\bibnamefont {Suenaga}},
  \bibinfo {author} {\bibfnamefont {S.}~\bibnamefont {Bandow}}, \bibinfo
  {author} {\bibfnamefont {H.}~\bibnamefont {Kato}}, \bibinfo {author}
  {\bibfnamefont {T.}~\bibnamefont {Okazaki}}, \bibinfo {author} {\bibfnamefont
  {H.}~\bibnamefont {Shinohara}}, \ and\ \bibinfo {author} {\bibfnamefont
  {S.}~\bibnamefont {Iijima}},\ }\href@noop {} {\bibfield  {journal} {\bibinfo
  {journal} {Phys. Rev. Lett.}\ }\textbf {\bibinfo {volume} {85}},\ \bibinfo
  {pages} {5384} (\bibinfo {year} {2000})}\BibitemShut {NoStop}%
\bibitem [{\citenamefont {Sun}\ \emph {et~al.}(2004)\citenamefont {Sun},
  \citenamefont {Inoue}, \citenamefont {Shimada}, \citenamefont {Okazaki},
  \citenamefont {Sugai}, \citenamefont {Suenaga},\ and\ \citenamefont
  {Shinohara}}]{sun2004synthesis}%
  \BibitemOpen
  \bibfield  {author} {\bibinfo {author} {\bibfnamefont {B.-Y.}\ \bibnamefont
  {Sun}}, \bibinfo {author} {\bibfnamefont {T.}~\bibnamefont {Inoue}}, \bibinfo
  {author} {\bibfnamefont {T.}~\bibnamefont {Shimada}}, \bibinfo {author}
  {\bibfnamefont {T.}~\bibnamefont {Okazaki}}, \bibinfo {author} {\bibfnamefont
  {T.}~\bibnamefont {Sugai}}, \bibinfo {author} {\bibfnamefont
  {K.}~\bibnamefont {Suenaga}}, \ and\ \bibinfo {author} {\bibfnamefont
  {H.}~\bibnamefont {Shinohara}},\ }\href@noop {} {\bibfield  {journal}
  {\bibinfo  {journal} {J. Phys. Chem. B}\ }\textbf {\bibinfo {volume} {108}},\
  \bibinfo {pages} {9011} (\bibinfo {year} {2004})}\BibitemShut {NoStop}%
\bibitem [{\citenamefont {Shiozawa}\ \emph {et~al.}(2006)\citenamefont
  {Shiozawa}, \citenamefont {Rauf}, \citenamefont {Pichler}, \citenamefont
  {Knupfer}, \citenamefont {Kalbac}, \citenamefont {Yang}, \citenamefont
  {Dunsch}, \citenamefont {B{\"u}chner}, \citenamefont {Batchelor},\ and\
  \citenamefont {Kataura}}]{shiozawa2006filling}%
  \BibitemOpen
  \bibfield  {author} {\bibinfo {author} {\bibfnamefont {H.}~\bibnamefont
  {Shiozawa}}, \bibinfo {author} {\bibfnamefont {H.}~\bibnamefont {Rauf}},
  \bibinfo {author} {\bibfnamefont {T.}~\bibnamefont {Pichler}}, \bibinfo
  {author} {\bibfnamefont {M.}~\bibnamefont {Knupfer}}, \bibinfo {author}
  {\bibfnamefont {M.}~\bibnamefont {Kalbac}}, \bibinfo {author} {\bibfnamefont
  {S.}~\bibnamefont {Yang}}, \bibinfo {author} {\bibfnamefont {L.}~\bibnamefont
  {Dunsch}}, \bibinfo {author} {\bibfnamefont {B.}~\bibnamefont {B{\"u}chner}},
  \bibinfo {author} {\bibfnamefont {D.}~\bibnamefont {Batchelor}}, \ and\
  \bibinfo {author} {\bibfnamefont {H.}~\bibnamefont {Kataura}},\ }\href@noop
  {} {\bibfield  {journal} {\bibinfo  {journal} {Phys. Rev. B}\ }\textbf
  {\bibinfo {volume} {73}},\ \bibinfo {pages} {205411} (\bibinfo {year}
  {2006})}\BibitemShut {NoStop}%
\bibitem [{\citenamefont {Kitaura}\ \emph {et~al.}(2007)\citenamefont
  {Kitaura}, \citenamefont {Okimoto}, \citenamefont {Shinohara}, \citenamefont
  {Nakamura},\ and\ \citenamefont {Osawa}}]{kitaura2007magnetism}%
  \BibitemOpen
  \bibfield  {author} {\bibinfo {author} {\bibfnamefont {R.}~\bibnamefont
  {Kitaura}}, \bibinfo {author} {\bibfnamefont {H.}~\bibnamefont {Okimoto}},
  \bibinfo {author} {\bibfnamefont {H.}~\bibnamefont {Shinohara}}, \bibinfo
  {author} {\bibfnamefont {T.}~\bibnamefont {Nakamura}}, \ and\ \bibinfo
  {author} {\bibfnamefont {H.}~\bibnamefont {Osawa}},\ }\href@noop {}
  {\bibfield  {journal} {\bibinfo  {journal} {Phys. Rev. B}\ }\textbf {\bibinfo
  {volume} {76}},\ \bibinfo {pages} {172409} (\bibinfo {year}
  {2007})}\BibitemShut {NoStop}%
\bibitem [{\citenamefont {Nicholls}\ \emph {et~al.}(2010)\citenamefont
  {Nicholls}, \citenamefont {Sader}, \citenamefont {Warner}, \citenamefont
  {Plant}, \citenamefont {Porfyrakis}, \citenamefont {Nellist}, \citenamefont
  {Briggs},\ and\ \citenamefont {Cockayne}}]{nicholls2010direct}%
  \BibitemOpen
  \bibfield  {author} {\bibinfo {author} {\bibfnamefont {R.~J.}\ \bibnamefont
  {Nicholls}}, \bibinfo {author} {\bibfnamefont {K.}~\bibnamefont {Sader}},
  \bibinfo {author} {\bibfnamefont {J.~H.}\ \bibnamefont {Warner}}, \bibinfo
  {author} {\bibfnamefont {S.~R.}\ \bibnamefont {Plant}}, \bibinfo {author}
  {\bibfnamefont {K.}~\bibnamefont {Porfyrakis}}, \bibinfo {author}
  {\bibfnamefont {P.~D.}\ \bibnamefont {Nellist}}, \bibinfo {author}
  {\bibfnamefont {G.~A.~D.}\ \bibnamefont {Briggs}}, \ and\ \bibinfo {author}
  {\bibfnamefont {D.~J.~H.}\ \bibnamefont {Cockayne}},\ }\href@noop {}
  {\bibfield  {journal} {\bibinfo  {journal} {ACS Nano}\ }\textbf {\bibinfo
  {volume} {4}},\ \bibinfo {pages} {3943} (\bibinfo {year} {2010})}\BibitemShut
  {NoStop}%
\bibitem [{\citenamefont {Yang}\ \emph {et~al.}(2010)\citenamefont {Yang},
  \citenamefont {Xu}, \citenamefont {Wei}, \citenamefont {Feng},\ and\
  \citenamefont {Suter}}]{yang2010quantum}%
  \BibitemOpen
  \bibfield  {author} {\bibinfo {author} {\bibfnamefont {W.~L.}\ \bibnamefont
  {Yang}}, \bibinfo {author} {\bibfnamefont {Z.~Y.}\ \bibnamefont {Xu}},
  \bibinfo {author} {\bibfnamefont {H.}~\bibnamefont {Wei}}, \bibinfo {author}
  {\bibfnamefont {M.}~\bibnamefont {Feng}}, \ and\ \bibinfo {author}
  {\bibfnamefont {D.}~\bibnamefont {Suter}},\ }\href@noop {} {\bibfield
  {journal} {\bibinfo  {journal} {Phys. Rev. A}\ }\textbf {\bibinfo {volume}
  {81}},\ \bibinfo {pages} {032303} (\bibinfo {year} {2010})}\BibitemShut
  {NoStop}%
\bibitem [{\citenamefont {Fritz}\ \emph {et~al.}(2017)\citenamefont {Fritz},
  \citenamefont {Westerstr{\"o}m}, \citenamefont {Kostanyan}, \citenamefont
  {Schlesier}, \citenamefont {Dreiser}, \citenamefont {Watts}, \citenamefont
  {Houben}, \citenamefont {Luysberg}, \citenamefont {Avdoshenko}, \citenamefont
  {Popov} \emph {et~al.}}]{fritz2017nanoscale}%
  \BibitemOpen
  \bibfield  {author} {\bibinfo {author} {\bibfnamefont {F.}~\bibnamefont
  {Fritz}}, \bibinfo {author} {\bibfnamefont {R.}~\bibnamefont
  {Westerstr{\"o}m}}, \bibinfo {author} {\bibfnamefont {A.}~\bibnamefont
  {Kostanyan}}, \bibinfo {author} {\bibfnamefont {C.}~\bibnamefont
  {Schlesier}}, \bibinfo {author} {\bibfnamefont {J.}~\bibnamefont {Dreiser}},
  \bibinfo {author} {\bibfnamefont {B.}~\bibnamefont {Watts}}, \bibinfo
  {author} {\bibfnamefont {L.}~\bibnamefont {Houben}}, \bibinfo {author}
  {\bibfnamefont {M.}~\bibnamefont {Luysberg}}, \bibinfo {author}
  {\bibfnamefont {S.~M.}\ \bibnamefont {Avdoshenko}}, \bibinfo {author}
  {\bibfnamefont {A.~A.}\ \bibnamefont {Popov}},  \emph {et~al.},\ }\href@noop
  {} {\bibfield  {journal} {\bibinfo  {journal} {Nanotechnology}\ }\textbf
  {\bibinfo {volume} {28}},\ \bibinfo {pages} {435703} (\bibinfo {year}
  {2017})}\BibitemShut {NoStop}%
\bibitem [{\citenamefont {Mazurenko}\ \emph {et~al.}(2017)\citenamefont
  {Mazurenko}, \citenamefont {Chiu}, \citenamefont {Ji}, \citenamefont
  {Parsons}, \citenamefont {Kan{\'a}sz-Nagy}, \citenamefont {Schmidt},
  \citenamefont {Grusdt}, \citenamefont {Demler}, \citenamefont {Greif},\ and\
  \citenamefont {Greiner}}]{mazurenko2017cold}%
  \BibitemOpen
  \bibfield  {author} {\bibinfo {author} {\bibfnamefont {A.}~\bibnamefont
  {Mazurenko}}, \bibinfo {author} {\bibfnamefont {C.~S.}\ \bibnamefont {Chiu}},
  \bibinfo {author} {\bibfnamefont {G.}~\bibnamefont {Ji}}, \bibinfo {author}
  {\bibfnamefont {M.~F.}\ \bibnamefont {Parsons}}, \bibinfo {author}
  {\bibfnamefont {M.}~\bibnamefont {Kan{\'a}sz-Nagy}}, \bibinfo {author}
  {\bibfnamefont {R.}~\bibnamefont {Schmidt}}, \bibinfo {author} {\bibfnamefont
  {F.}~\bibnamefont {Grusdt}}, \bibinfo {author} {\bibfnamefont
  {E.}~\bibnamefont {Demler}}, \bibinfo {author} {\bibfnamefont
  {D.}~\bibnamefont {Greif}}, \ and\ \bibinfo {author} {\bibfnamefont
  {M.}~\bibnamefont {Greiner}},\ }\href@noop {} {\bibfield  {journal} {\bibinfo
   {journal} {Nature}\ }\textbf {\bibinfo {volume} {545}},\ \bibinfo {pages}
  {462} (\bibinfo {year} {2017})}\BibitemShut {NoStop}%
\bibitem [{\citenamefont {Tai}\ \emph {et~al.}(2017)\citenamefont {Tai},
  \citenamefont {Lukin}, \citenamefont {Rispoli}, \citenamefont {Schittko},
  \citenamefont {Menke}, \citenamefont {Borgnia}, \citenamefont {Preiss},
  \citenamefont {Grusdt}, \citenamefont {Kaufman},\ and\ \citenamefont
  {Greiner}}]{tai2017microscopy}%
  \BibitemOpen
  \bibfield  {author} {\bibinfo {author} {\bibfnamefont {M.~E.}\ \bibnamefont
  {Tai}}, \bibinfo {author} {\bibfnamefont {A.}~\bibnamefont {Lukin}}, \bibinfo
  {author} {\bibfnamefont {M.}~\bibnamefont {Rispoli}}, \bibinfo {author}
  {\bibfnamefont {R.}~\bibnamefont {Schittko}}, \bibinfo {author}
  {\bibfnamefont {T.}~\bibnamefont {Menke}}, \bibinfo {author} {\bibfnamefont
  {D.}~\bibnamefont {Borgnia}}, \bibinfo {author} {\bibfnamefont {P.~M.}\
  \bibnamefont {Preiss}}, \bibinfo {author} {\bibfnamefont {F.}~\bibnamefont
  {Grusdt}}, \bibinfo {author} {\bibfnamefont {A.~M.}\ \bibnamefont {Kaufman}},
  \ and\ \bibinfo {author} {\bibfnamefont {M.}~\bibnamefont {Greiner}},\
  }\href@noop {} {\bibfield  {journal} {\bibinfo  {journal} {Nature}\ }\textbf
  {\bibinfo {volume} {546}},\ \bibinfo {pages} {519} (\bibinfo {year}
  {2017})}\BibitemShut {NoStop}%
\bibitem [{\citenamefont {Gorshkov}\ \emph {et~al.}(2011)\citenamefont
  {Gorshkov}, \citenamefont {Manmana}, \citenamefont {Chen}, \citenamefont
  {Ye}, \citenamefont {Demler}, \citenamefont {Lukin},\ and\ \citenamefont
  {Rey}}]{gorshkov2011tunable}%
  \BibitemOpen
  \bibfield  {author} {\bibinfo {author} {\bibfnamefont {A.~V.}\ \bibnamefont
  {Gorshkov}}, \bibinfo {author} {\bibfnamefont {S.~R.}\ \bibnamefont
  {Manmana}}, \bibinfo {author} {\bibfnamefont {G.}~\bibnamefont {Chen}},
  \bibinfo {author} {\bibfnamefont {J.}~\bibnamefont {Ye}}, \bibinfo {author}
  {\bibfnamefont {E.}~\bibnamefont {Demler}}, \bibinfo {author} {\bibfnamefont
  {M.~D.}\ \bibnamefont {Lukin}}, \ and\ \bibinfo {author} {\bibfnamefont
  {A.~M.}\ \bibnamefont {Rey}},\ }\href@noop {} {\bibfield  {journal} {\bibinfo
   {journal} {Phys. Rev. Lett.}\ }\textbf {\bibinfo {volume} {107}},\ \bibinfo
  {pages} {115301} (\bibinfo {year} {2011})}\BibitemShut {NoStop}%
\bibitem [{\citenamefont {Abolins}, \citenamefont {Zillich},\ and\
  \citenamefont {Whaley}(2011)}]{abolins2011ground}%
  \BibitemOpen
  \bibfield  {author} {\bibinfo {author} {\bibfnamefont {B.~P.}\ \bibnamefont
  {Abolins}}, \bibinfo {author} {\bibfnamefont {R.~E.}\ \bibnamefont
  {Zillich}}, \ and\ \bibinfo {author} {\bibfnamefont {K.~B.}\ \bibnamefont
  {Whaley}},\ }\href@noop {} {\bibfield  {journal} {\bibinfo  {journal} {J. Low
  Temp. Phys.}\ }\textbf {\bibinfo {volume} {165}},\ \bibinfo {pages} {249}
  (\bibinfo {year} {2011})}\BibitemShut {NoStop}%
\bibitem [{\citenamefont {White}(1992)}]{white1992density}%
  \BibitemOpen
  \bibfield  {author} {\bibinfo {author} {\bibfnamefont {S.~R.}\ \bibnamefont
  {White}},\ }\href@noop {} {\bibfield  {journal} {\bibinfo  {journal} {Phys.
  Rev. Lett.}\ }\textbf {\bibinfo {volume} {69}},\ \bibinfo {pages} {2863}
  (\bibinfo {year} {1992})}\BibitemShut {NoStop}%
\bibitem [{\citenamefont {Schollw{\"o}ck}(2005)}]{schollwock2005density}%
  \BibitemOpen
  \bibfield  {author} {\bibinfo {author} {\bibfnamefont {U.}~\bibnamefont
  {Schollw{\"o}ck}},\ }\href@noop {} {\bibfield  {journal} {\bibinfo  {journal}
  {Rev. Mod. Phys.}\ }\textbf {\bibinfo {volume} {77}},\ \bibinfo {pages} {259}
  (\bibinfo {year} {2005})}\BibitemShut {NoStop}%
\bibitem [{\citenamefont {Schollw{\"o}ck}(2011)}]{schollwock2011density}%
  \BibitemOpen
  \bibfield  {author} {\bibinfo {author} {\bibfnamefont {U.}~\bibnamefont
  {Schollw{\"o}ck}},\ }\href@noop {} {\bibfield  {journal} {\bibinfo  {journal}
  {Ann. Phys.}\ }\textbf {\bibinfo {volume} {326}},\ \bibinfo {pages} {96}
  (\bibinfo {year} {2011})}\BibitemShut {NoStop}%
\bibitem [{\citenamefont {White}\ and\ \citenamefont
  {Martin}(1999)}]{white1999ab}%
  \BibitemOpen
  \bibfield  {author} {\bibinfo {author} {\bibfnamefont {S.~R.}\ \bibnamefont
  {White}}\ and\ \bibinfo {author} {\bibfnamefont {R.~L.}\ \bibnamefont
  {Martin}},\ }\href@noop {} {\bibfield  {journal} {\bibinfo  {journal} {J.
  Chem. Phys.}\ }\textbf {\bibinfo {volume} {110}},\ \bibinfo {pages} {4127}
  (\bibinfo {year} {1999})}\BibitemShut {NoStop}%
\bibitem [{\citenamefont {Chan}\ and\ \citenamefont
  {Sharma}(2011)}]{chan2011density}%
  \BibitemOpen
  \bibfield  {author} {\bibinfo {author} {\bibfnamefont {G.~K.-L.}\
  \bibnamefont {Chan}}\ and\ \bibinfo {author} {\bibfnamefont {S.}~\bibnamefont
  {Sharma}},\ }\href@noop {} {\bibfield  {journal} {\bibinfo  {journal} {Annu.
  Rev. Phys. Chem.}\ }\textbf {\bibinfo {volume} {62}},\ \bibinfo {pages} {465}
  (\bibinfo {year} {2011})}\BibitemShut {NoStop}%
\bibitem [{\citenamefont {Duperrouzel}\ \emph {et~al.}(2015)\citenamefont
  {Duperrouzel}, \citenamefont {Tecmer}, \citenamefont {Boguslawski},
  \citenamefont {Barcza}, \citenamefont {Legeza},\ and\ \citenamefont
  {Ayers}}]{duperrouzel2015quantum}%
  \BibitemOpen
  \bibfield  {author} {\bibinfo {author} {\bibfnamefont {C.}~\bibnamefont
  {Duperrouzel}}, \bibinfo {author} {\bibfnamefont {P.}~\bibnamefont {Tecmer}},
  \bibinfo {author} {\bibfnamefont {K.}~\bibnamefont {Boguslawski}}, \bibinfo
  {author} {\bibfnamefont {G.}~\bibnamefont {Barcza}}, \bibinfo {author}
  {\bibfnamefont {{\"O}.}~\bibnamefont {Legeza}}, \ and\ \bibinfo {author}
  {\bibfnamefont {P.~W.}\ \bibnamefont {Ayers}},\ }\href@noop {} {\bibfield
  {journal} {\bibinfo  {journal} {Chem. Phys. Lett.}\ }\textbf {\bibinfo
  {volume} {621}},\ \bibinfo {pages} {160} (\bibinfo {year}
  {2015})}\BibitemShut {NoStop}%
\bibitem [{\citenamefont {Stoudenmire}\ and\ \citenamefont
  {White}(2012)}]{stoudenmire2012studying}%
  \BibitemOpen
  \bibfield  {author} {\bibinfo {author} {\bibfnamefont {E.~M.}\ \bibnamefont
  {Stoudenmire}}\ and\ \bibinfo {author} {\bibfnamefont {S.~R.}\ \bibnamefont
  {White}},\ }\href@noop {} {\bibfield  {journal} {\bibinfo  {journal} {Annu.
  Rev. Condens. Matter Phys.}\ }\textbf {\bibinfo {volume} {3}},\ \bibinfo
  {pages} {111} (\bibinfo {year} {2012})}\BibitemShut {NoStop}%
\bibitem [{ite()}]{itensor}%
  \BibitemOpen
  \href@noop {} {\enquote {\bibinfo {title} {{ITensor} library (version 2.1.1)
  http://itensor.org.}}\ }\BibitemShut {NoStop}%
\bibitem [{\citenamefont {Ruhman}\ \emph {et~al.}(2012)\citenamefont {Ruhman},
  \citenamefont {Dalla~Torre}, \citenamefont {Huber},\ and\ \citenamefont
  {Altman}}]{ruhman2012nonlocal}%
  \BibitemOpen
  \bibfield  {author} {\bibinfo {author} {\bibfnamefont {J.}~\bibnamefont
  {Ruhman}}, \bibinfo {author} {\bibfnamefont {E.~G.}\ \bibnamefont
  {Dalla~Torre}}, \bibinfo {author} {\bibfnamefont {S.~D.}\ \bibnamefont
  {Huber}}, \ and\ \bibinfo {author} {\bibfnamefont {E.}~\bibnamefont
  {Altman}},\ }\href@noop {} {\bibfield  {journal} {\bibinfo  {journal} {Phys.
  Rev. B}\ }\textbf {\bibinfo {volume} {85}},\ \bibinfo {pages} {125121}
  (\bibinfo {year} {2012})}\BibitemShut {NoStop}%
\bibitem [{\citenamefont {Manmana}\ \emph {et~al.}(2013)\citenamefont
  {Manmana}, \citenamefont {Stoudenmire}, \citenamefont {Hazzard},
  \citenamefont {Rey},\ and\ \citenamefont
  {Gorshkov}}]{manmana2013topological}%
  \BibitemOpen
  \bibfield  {author} {\bibinfo {author} {\bibfnamefont {S.~R.}\ \bibnamefont
  {Manmana}}, \bibinfo {author} {\bibfnamefont {E.~M.}\ \bibnamefont
  {Stoudenmire}}, \bibinfo {author} {\bibfnamefont {K.~R.~A.}\ \bibnamefont
  {Hazzard}}, \bibinfo {author} {\bibfnamefont {A.~M.}\ \bibnamefont {Rey}}, \
  and\ \bibinfo {author} {\bibfnamefont {A.~V.}\ \bibnamefont {Gorshkov}},\
  }\href@noop {} {\bibfield  {journal} {\bibinfo  {journal} {Phys. Rev. B}\
  }\textbf {\bibinfo {volume} {87}},\ \bibinfo {pages} {081106} (\bibinfo
  {year} {2013})}\BibitemShut {NoStop}%
\bibitem [{\citenamefont {Ballentine}(1990)}]{ballentine1990quantum}%
  \BibitemOpen
  \bibfield  {author} {\bibinfo {author} {\bibfnamefont {L.~E.}\ \bibnamefont
  {Ballentine}},\ }\href@noop {} {\emph {\bibinfo {title} {Quantum
  Mechanics}}}\ (\bibinfo  {publisher} {Prentice-Hall},\ \bibinfo {year}
  {1990})\BibitemShut {NoStop}%
\bibitem [{\citenamefont {Liu}, \citenamefont {Xun},\ and\ \citenamefont
  {Shan}(2010)}]{liu2010raising}%
  \BibitemOpen
  \bibfield  {author} {\bibinfo {author} {\bibfnamefont {Q.~H.}\ \bibnamefont
  {Liu}}, \bibinfo {author} {\bibfnamefont {D.~M.}\ \bibnamefont {Xun}}, \ and\
  \bibinfo {author} {\bibfnamefont {L.}~\bibnamefont {Shan}},\ }\href@noop {}
  {\bibfield  {journal} {\bibinfo  {journal} {Int. J. Theor. Phys.}\ }\textbf
  {\bibinfo {volume} {49}},\ \bibinfo {pages} {2164} (\bibinfo {year}
  {2010})}\BibitemShut {NoStop}%
\bibitem [{\citenamefont {Sachdev}(2011)}]{sachdev2011quantum}%
  \BibitemOpen
  \bibfield  {author} {\bibinfo {author} {\bibfnamefont {S.}~\bibnamefont
  {Sachdev}},\ }\href@noop {} {\emph {\bibinfo {title} {Quantum Phase
  Transitions}}},\ \bibinfo {edition} {2nd}\ ed.\ (\bibinfo  {publisher}
  {Cambridge University Press},\ \bibinfo {year} {2011})\BibitemShut {NoStop}%
\bibitem [{\citenamefont {Cardy}(1996)}]{cardy2002scaling}%
  \BibitemOpen
  \bibfield  {author} {\bibinfo {author} {\bibfnamefont {J.}~\bibnamefont
  {Cardy}},\ }\href@noop {} {\emph {\bibinfo {title} {Scaling and
  Renormalization in Statistical Physics}}}\ (\bibinfo  {publisher} {Cambridge
  University Press},\ \bibinfo {year} {1996})\BibitemShut {NoStop}%
\bibitem [{\citenamefont {Calabrese}\ and\ \citenamefont
  {Cardy}(2009)}]{calabrese2009entanglement}%
  \BibitemOpen
  \bibfield  {author} {\bibinfo {author} {\bibfnamefont {P.}~\bibnamefont
  {Calabrese}}\ and\ \bibinfo {author} {\bibfnamefont {J.}~\bibnamefont
  {Cardy}},\ }\href@noop {} {\bibfield  {journal} {\bibinfo  {journal} {J.
  Phys. A: Math. Theor.}\ }\textbf {\bibinfo {volume} {42}},\ \bibinfo {pages}
  {504005} (\bibinfo {year} {2009})}\BibitemShut {NoStop}%
\end{thebibliography}%

\end{document}